\documentclass[prd,aps,floats,epsfig,twocolumn]{revtex4}
\usepackage{graphicx}
\usepackage{color}
\newcommand{\beq}{\begin{equation}}
\newcommand{\eeq}{\end{equation}}
\newcommand{\bea}{\begin{eqnarray}}
\newcommand{\eea}{\end{eqnarray}}

\newcommand{\ApJ}{{\it Astrophys. J.\,}}

\newcommand{\PR}{{\it Phys. Rev.\,}}

\begin{document}
\title{Delayed Recombination and Cosmic Parameters}
\author{Silvia Galli$^{*}$, Rachel Bean$^\sharp$, Alessandro Melchiorri$^{*}$ and Joseph Silk $^\flat$}
\affiliation{$^*$ Universita' di Roma ``La Sapienza'', Ple Aldo Moro 2, 00185, Rome, Italy.\\
$^\sharp$ Dept. of Astronomy, Space Sciences Building, Cornell University, Ithaca, NY 14853, USA.\\
$^\flat$ Astrophysics, Denys Wilkinson Building, University of Oxford, Keble Road, OX1 3RH, Oxford, UK.}
\begin{abstract}

Current cosmological constraints from Cosmic Microwave Background (CMB)
anisotropies are typically derived assuming a standard
recombination scheme, however additional resonance and ionizing radiation sources 
can delay recombination, altering the cosmic ionization history and
the cosmological 
inferences drawn from CMB data. 
We show that for recent observations of CMB anisotropy, from 
the Wilkinson Microwave Anisotropy Probe satellite mission 5-year
survey (WMAP5) and from the ACBAR experiment,  additional resonance
radiation is nearly 
degenerate with
variations in the spectral index, $n_s$, and 
has a marked effect on uncertainties in constraints on the Hubble constant, age of the universe,
curvature and the upper bound on the neutrino mass. 
When a modified recombination scheme is considered, the
redshift of recombination is constrained to 
$z_*=1078\pm11$, with uncertainties in the measurement weaker by one order of magnitude than those obtained under the assumption of standard recombination
while constraints on the shift parameter are shifted by 1$\sigma$ to ${\cal R}=1.734\pm0.028$.
Although delayed recombination limits the precision of parameter
estimation from the WMAP satellite, we demonstrate that this should not be the case for
future, smaller angular scales measurements, such as those 
by the Planck satellite mission. 

\end{abstract}
\maketitle
\section{Introduction}
The recent measurements of the Cosmic Microwave 
Background (CMB) flux provided 
by the five year Wilkinson Microwave Anisotropy Probe (WMAP) 
mission (see \cite{wmap5cosm,wmap5komatsu}
and the ACBAR collaboration (see \cite{acbar})
have confirmed several aspects of the
cosmological standard model and improved the constraints
on several key parameters. The constraints on the neutrino
mass, for example, coming just from CMB data are 
drastically improved and are now competitive with current
direct laboratory measurements (see e.g. \cite{wmap5komatsu,fogli}). 
Moreover, the presence of a neutrino background is now inferred
at more than $95 \%$ c.l. by CMB data alone
(see e.g., \cite{wmap5komatsu,fdebe08}).
Finally, inflationary parameters, curvature, baryonic and dark matter 
densities are also now better determined
due to improved treatment of systematics
(see \cite{wmap5komatsu,kinney08}).

These spectacular results, apart from the experimental improvements, 
have been possible  due to the high precision of the 
CMB theoretical predictions that have now reached
an accuracy close to $0.1\%$ over a wide range 
of scales. 
A key ingredient in the CMB precision cosmology is 
the accurate computation of the recombination process.
Since the seminal papers by Peebles and Z'eldovich 
(see \cite{Peebles68,Zeldovich68}) detailing the 
recombination process, further refinements to the standard scheme 
were developed \cite{Seager99} allowing predictions at the accuracy level 
found in data from the WMAP satellite and predicted for 
the future Planck satellite \cite{Hu95,Seljak03,hirata}.  

While the attained accuracy on the recombination process is
impressive, it should be noticed that these computations rely on the
assumption of standard physics. Non-standard mechanisms 
such as (just to name a few) high redshift stars or 
active galactic nuclei, topological defects and
dark matter decays could produce extra sources of radiation and 
modify the recombination process. With the WMAP results and the future
Planck data, it therefore becomes conceivable that deviations from
standard recombination may be be detected.

Several papers in recent years have indeed investigated this possibility.
For example, one could use a phenomenological approach 
such as in \cite{Hannestad01} or \cite{Lewis:2006ym}. 
Other works have adopted a more physically motivated (but
model-dependent) approach and considered modified recombination 
by allowing time variations in the fine-structure (\cite{alpha})
or gravitational constants (\cite{zalzahn}).
Here we instead focus on delayed recombination mechanisms based on 
the hypothesis of extra sources of ionizing and resonance radiation 
at recombination (see e.g. \cite{Seager00,Naselsky02,Dorosh02}).

Extra photon sources can be generated by a variety of mechanisms. 
A widely considered process is the decay or annihilation of massive particles 
\cite{Sarkar:1983,Scott:1991,Ellis:1992,Adams:1998,Doroshkevich:2002ff,Naselsky:2003zj,Zhang:2006fr}.  
The decay channel depends on the nature of the particles, and could,
for example, include charged and neutral leptons, quarks or gauge 
bosons.  These particles may then decay further, leading to a shower cascade
that could,  among other products, generate a bath of lower energy photons that 
would interact with the primordial gas and cosmic microwave
background.  
Interestingly these models, as well as injecting energy at
recombination, boost the ionization fraction after recombination and 
can distort the ionization history of the universe at even later
times, during galaxy formation and reionization 
\cite{Pierpaoli:2003rz,Chen:2003gz,Padmanabhan:2005es,Mapelli:2006ej,Lewis:2006ma}. 
Other mechanisms include evaporation of black holes
\cite{Naselsky87,Naselsky02} or inhomogenities in baryonic matter \cite{Naselsky02}.

Several authors have already compared 
delayed recombination with cosmological data (see for example 
\cite{bms,bms2,naso,petruta,cinesi}).
With respect to these previous analyses, we assess below the improvements 
given by more recent data from the WMAP five year survey and 
from the ACBAR experiment. Moreover, we will study in detail the
impact of delayed recombination on the current constraints
on cosmic parameters. As showed in \cite{Seager00} delayed
recombination has two main effects: damping of the
CMB anisotropy and polarization at small angular scales
and a shift of the acoustic peaks in their angular
spectra. As we will see, these two effects change in a significant
way the current constraints on such parameters as the spectral index $n_S$ and
the Hubble constant $H_0$. We then study the impact 
on the current determination of the redshift of recombination $z_*$ and
on the shift parameter $\cal R$. Both parameters are used to provide
complementary 
geometric constraints on the background expansion history; $R$ as a
measure of the 
angular diameter distance to last scattering, and $z_*$ in
interpreting the scale of baryon 
acoustic oscillations \cite{Eisenstein:2005su,Percival:2007yw}, which in combination with supernova 
type Ia data, can impose constraints on the dark energy
component (see e.g. \cite{efstathiou,corasaniti,Elgaroy:2007bv}. Here we show that delayed recombination has
decisive effects on the determination of these parameters.
Finally, we  discuss the implications of a modified recombination
scheme for the Planck satellite. 

The paper proceeds as follows: in section \ref{sec2} we briefly
describe the delayed recombination scheme.
In \ref{sec3} we analyze the latest CMB data and place new 
constraints on delayed recombination. Moreover, we forecast 
the constraints prospectively attainable by the Planck satellite mission, and
 study the impact that a modified recombination 
scheme can have on several key cosmological and astrophysical parameters.
In \ref{sec4} we draw our conclusions and review their future implications.

\section{A modified ionization history}\label{sec2}

Following \cite{Peebles68,Zeldovich68} we can model the evolution of the electron ionization fraction, $x_{e}$
in a simplified manner for the
recombination of hydrogen:
\bea 
-{dx_{e}\over dt}\left.\right|_{std}=C\left[a_c n x_{e}^{2}-b_c 
(1-x_{e})\exp{\left(-{\Delta B\over k_{B}T}\right)}\right]
\label{eq1}
\eea
where $n$ is the number density of atoms, $a_{c}$ and $b_{c}$ are 
the effective recombination and photo-ionization rates for principle 
quantum numbers $\ge 2$, $\Delta B$ is the difference in binding energy between the $1^{st}$ and $2^{nd}$ energy levels and
\bea C={1+K\Lambda_{1s2s}n_{1s}\over1+K(\Lambda_{1s2s}+b_{c})n_{1s}}
\label{eqC},  \ \ \ \ K={\lambda_\alpha^{3} \over8\pi H(z)}
\eea
where  $\lambda_{\alpha}$ is the wavelength of the single Ly-$\alpha$
transition from the $2p$ level, $\Lambda_{1s2s}$ is 
the decay rate of the metastable $2s$ level, $n_{1s}=n(1-x_{e})$ is
the number of neutral ground state $H$ atoms, and $H(z)$ 
is the Hubble expansion factor at a redshift $z$.

As in \cite{bms2}, we include the possibility of extra photons at key wavelengths that
could modify this recombination picture, namely resonance
(Ly-$\alpha$) 
photons with number density, $n_{\alpha}$,which promote electrons to
the $2p$ level, and ionizing photons, 
$n_{i}$,\cite{Seager00,Naselsky02,Dorosh02,bms}
\bea
{dn_{\alpha}\over dt}&=&\varepsilon_{\alpha}(z)H(z)n, \ \ \ 
{dn_{i}\over dt}=\varepsilon_{i}(z)H(z)n.
\label{eq0}
\eea
This leads to a modified evolution of the ionization fraction
\bea -{dx_{e}\over dt}&=&-{dx_{e}\over dt}\left.
\right|_{std}-C\varepsilon_{i} H-(1-C)\varepsilon_{\alpha}H . 
\ \ \ \ \ \ \ \label{eq2}
\eea

We employ the widely used \texttt{RECFAST} code \cite{Seager99}, in
the \texttt{cosmomc} 
package \cite{Lewis:2002ah} modifying the code as in (\ref{eq2}) to
include two extra 
constant parameters, $\epsilon_{\alpha}$ and $\epsilon_{i}$. 

\begin{figure}[t]
\begin{center}
\includegraphics[width=3.5in]{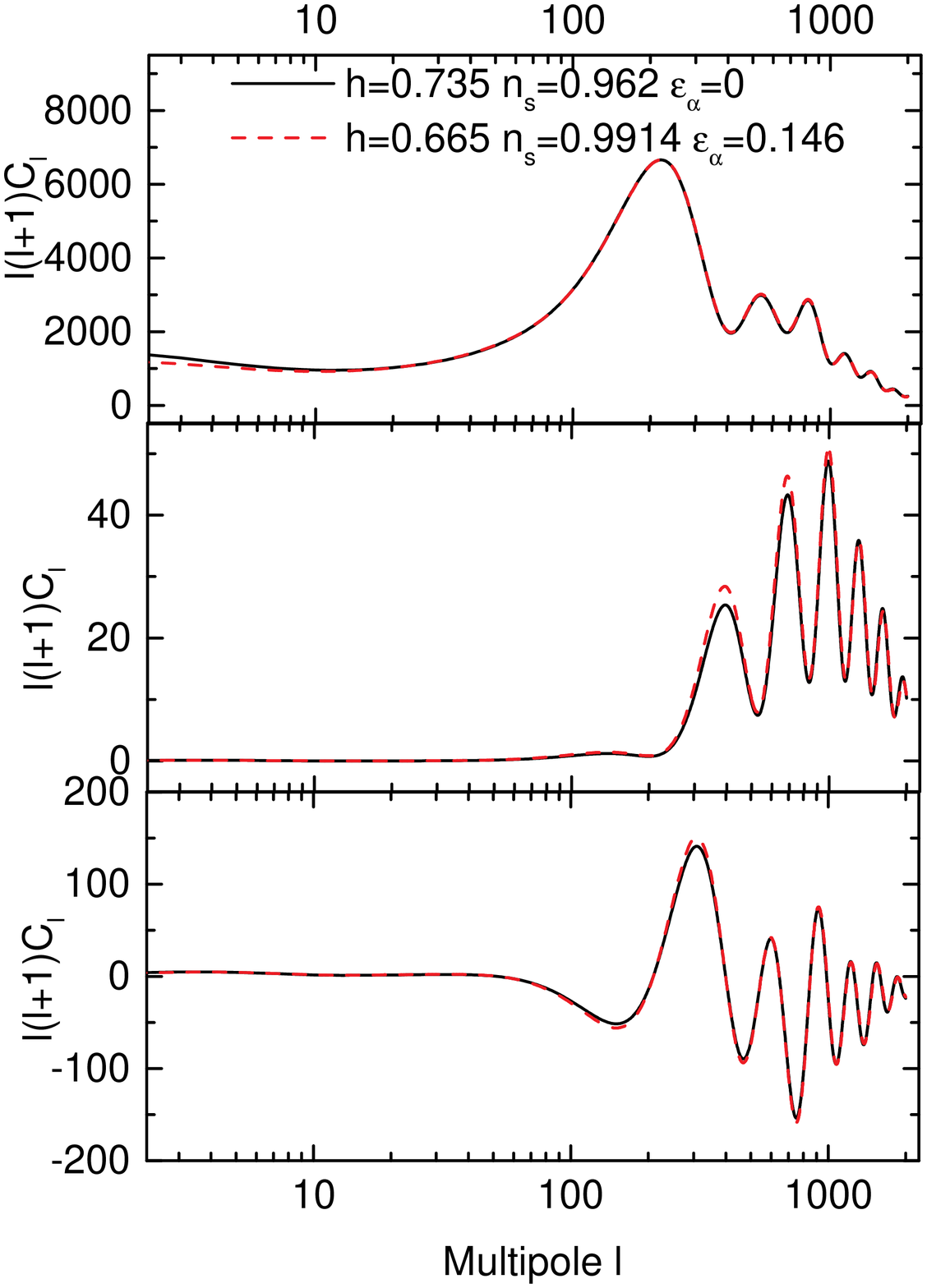}
\caption{Best fit WMAP5 Spectra and degenerate 
Anisotropy (Top), Polarization (Middle Panel) and cross
Anisotropy-Polarization (Bottom Panel) angular power spectra in case
of delayed recombination.}
\label{fig0}
\end{center}
\end{figure}

The effects of delayed recombination on the primordial power
spectra have been througly investigated by \cite{Seager00}.
Namely, the main consequences are damping of CMB anisotropies and
polarization on small angular scales and a shift of 
the acoustic peaks towards larger scales. This introduces a
main degeneracy with two parameters: the scalar spectral index
$n_S$ and, in a flat universe, the Hubble constant $H_0$.
In Figure \ref{fig0} we indeed plot anisotropy, 
polarization and cross temperature-anisotropy power spectra
for two degenerate models. As we see, when delayed recombination
is included, the degenerate spectra have a larger spectal index
to compensate the damping and a smaller Hubble parameter
to compensate the shift. This degeneracy will clearly appear
as a correlation between the constraints on these parameters
as we show in the next section. 

We compare delayed recombination with current data by making use
of the the publicly available Markov Chain Monte Carlo
package \texttt{cosmomc} \cite{Lewis:2002ah}. 
Other than $\varepsilon_{\alpha}$ and $\varepsilon_{i}$,
we sample the following
seven-dimensional set of cosmological parameters, adopting flat priors on them:
the physical baryon and CDM densities, $\omega_b=\Omega_bh^2$ and
$\omega_c=\Omega_ch^2$, the Hubble parameter $H_0$, the scalar
spectral index, $n_{s}$, 
the normalization, $\ln10^{10}A_s(k=0.05/Mpc)$,
and the optical depth to reionization, $\tau$. 
We also consider the
possibility of having a curved universe with 
$\omega_k\neq0$ \footnote{In the case of curved models 
we consider the ratio of the sound horizon and angular distance 
at recombinationparameter $\theta$ instead of $H_0$, modified accordingly
to \cite{Seager00}} and
three massive neutrinos 
each with same mass $m_{\nu}$ and overall energy density:

\bea
\omega_{\nu}=\frac{m_{\nu}}{30.8 eV}
\eea

We consider purely adiabatic initial conditions.
In what follows we also consider bounds on the derived
``shift'' parameter $\cal R$ defined as (see \cite{efstathiou}):

\begin{equation}
{\cal R}=\frac{\omega_{\rm m}^{1/2}}{\omega_{k}^{1/2}}{\rm sinn}_k
(\omega_{k}^{1/2}y),
\label{ebr}
\end{equation}
where $\textrm{sinn}(x)=\{\sin(x),x,\sinh(x)\}$ for closed, flat and 
open geometries respectively, with 
\begin{equation}
y=\int_{a_{\rm *}}^1\frac{da}{\sqrt{\omega_{\rm m} a+\omega_{\rm k} a^2+
\omega_\Lambda a^4+\omega_{\Lambda}}}
\label{ebry}
\end{equation}

\noindent where $\omega_m=\omega_b+\omega_c$ and $a$ is the scale 
factor with $a_*=(1+z_*)^{-1}$.
The MCMC convergence diagnostic tests are performed on $8$ chains using the
Gelman and Rubin ``variance of chain mean''$/$``mean of chain variances'' $R$
statistic for each parameter. Our $1-D$ and $2-D$ constraints are obtained
after marginalization over the remaining ``nuisance'' parameters, again using
the programs included in the \texttt{cosmomc} package. 
We use a cosmic age top-hat prior as 10 Gyr $ \le t_0 \le$ 20 Gyr.
We include the five-year WMAP data \cite{wmap5komatsu} (temperature
and polarization) with the routine for computing the likelihood
supplied by the WMAP team (we will refer to this analysis as WMAP5). 
Together with the WMAP data we also
consider the small-scale CMB measurements of ACBAR
\cite{acbar} (we will refer to this analysis as WMAP5+ACBAR).

\section{Likelihood analysis and Results}
\label{sec3}

\subsection{Modified Recombination with standard $\Lambda$-CDM}
We first analyze the effects of modified recombination
on the cosmological constraints obtained under
the assumption of a flat, $\Lambda$-CDM model
with massless neutrinos.
After marginalization over the 
nuisance parameters described in the previous section,
the WMAP5 data gives
$\epsilon_\alpha < 0.39$ and $\epsilon_i < 0.058$
at $95 \%$ c.l. ($\epsilon_\alpha < 0.31$ and 
$\epsilon_i < 0.053$ when
the ACBAR data is also included).

\begin{figure}[t]
\begin{center}
\includegraphics[width=2.5in]{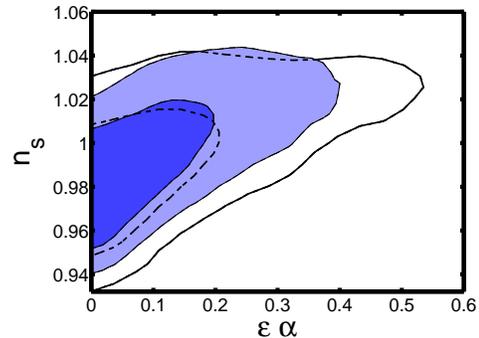}
\caption{Likelihood contours at  $68 \%$ and 
$95 \%$ c.l. in the $n_s-\epsilon_\alpha$,
plane for WMAP5 (empty contours) and
WMAP5+ACBAR (filled contours) experiments respectively, allowing both ionizing and resonance radiation modifications to the recombination scheme.
}
\label{fig1}
\end{center}
\end{figure}

\begin{figure}[t]
\begin{center}
\includegraphics[width=2.5in]{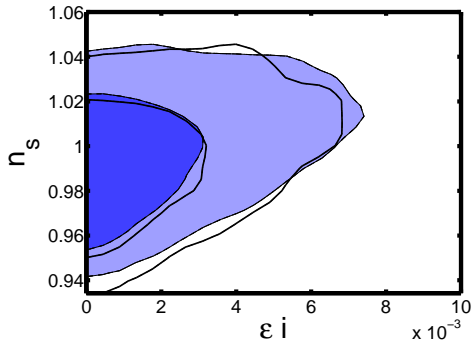}
\caption{Likelihood contours at  $68 \%$ and 
$95 \%$ c.l. in the $n_s-\epsilon_i$,
plane for WMAP5 (empty contours) and
WMAP5+ACBAR (filled contours) experiments respectively, allowing both ionizing and resonance radiation modifications to the recombination scheme.
}
\label{fig2}
\end{center}
\end{figure}

In Figure \ref{fig1} and Figure \ref{fig2} 
we plot the $68 \%$ and $95 \%$ c.l. likelihood contours 
on the $n_s-\epsilon_\alpha$,
and  $n_s-\epsilon_i$ planes for the WMAP5 and
WMAP5+ACBAR datasets respectively.
The figures highlight that even with the improved small
scale CMB data, modified 
recombination's main effect is to drive $n_s$ to higher values, reconciling the
data with a $n_s=1$, Harrison-Z'eldovich (HZ), spectrum, as was seen with the WMAP 3-year data \cite{bms2}. 
Any conclusion on the compatibility of a particular inflationary
model with the WMAP5 data, therefore, is highly dependent on the
assumptions made on the recombination process and should
 be discussed with some caution in light of this.

Marginalizing over the recombination parameters we 
indeed get 
$n_s=0.993_{-0.040}^{+0.045}$ (WMAP5 alone) and 
$n_s=0.996_{-0.038}^{+0.042}$ (WMAP5+ACBAR) 
at $95 \%$ c.l.., i.e. with $n_s=1$ 
perfectly compatible with the data.
Those results should be compared with the constraints
 $n_s=0.959_{-0.027}^{+0.026}$ and 
$n_s=0.965_{-0.028}^{+0.027}$ ($95 \%$ c.l.) obtained using the same
datasets and priors but with standard recombination.

\begin{figure}[t]
\begin{center}
\includegraphics[width=2.5in]{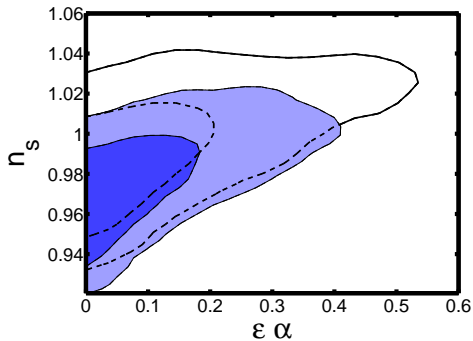}
\caption{Likelihood contours at  $68 \%$ and 
$95 \%$ c.l. on the $n_s-\epsilon_\alpha$,
plane for WMAP5 assuming variations in 
 $\epsilon_i$ (empty contours) and
fixing  $\epsilon_i=0$ (filled contours).}
\label{fig3}
\end{center}
\end{figure}

As we can see from Figure \ref{fig1} it is possible
to have $n_s=1$ even if $\epsilon_{\alpha}=0$.
However this agreement, apparently in contrast with
the standard WMAP5 result, is due to the marginalization
over $\epsilon_i$ which is still present as a free parameter 
in the analysis. Since it may be possible 
to have only resonance radiation at recombination 
we analyze the effects of just varying $\epsilon_\alpha$
while keeping  $\epsilon_i=0$. In Figure \ref{fig3} we plot the
constraints on the $n_s$-$\epsilon_{\alpha}$ plane with and without
variation in $\epsilon_i$. As we can see, when $\epsilon_i=0$,
the degeneracy between $n_s$ and $\epsilon_{\alpha}$ is more 
evident and HZ spectra are in agreement at $1$-$\sigma$ level with
WMAP5 only if delayed recombination is present.

\begin{figure}[t]
\begin{center}
\includegraphics[width=2.5in]{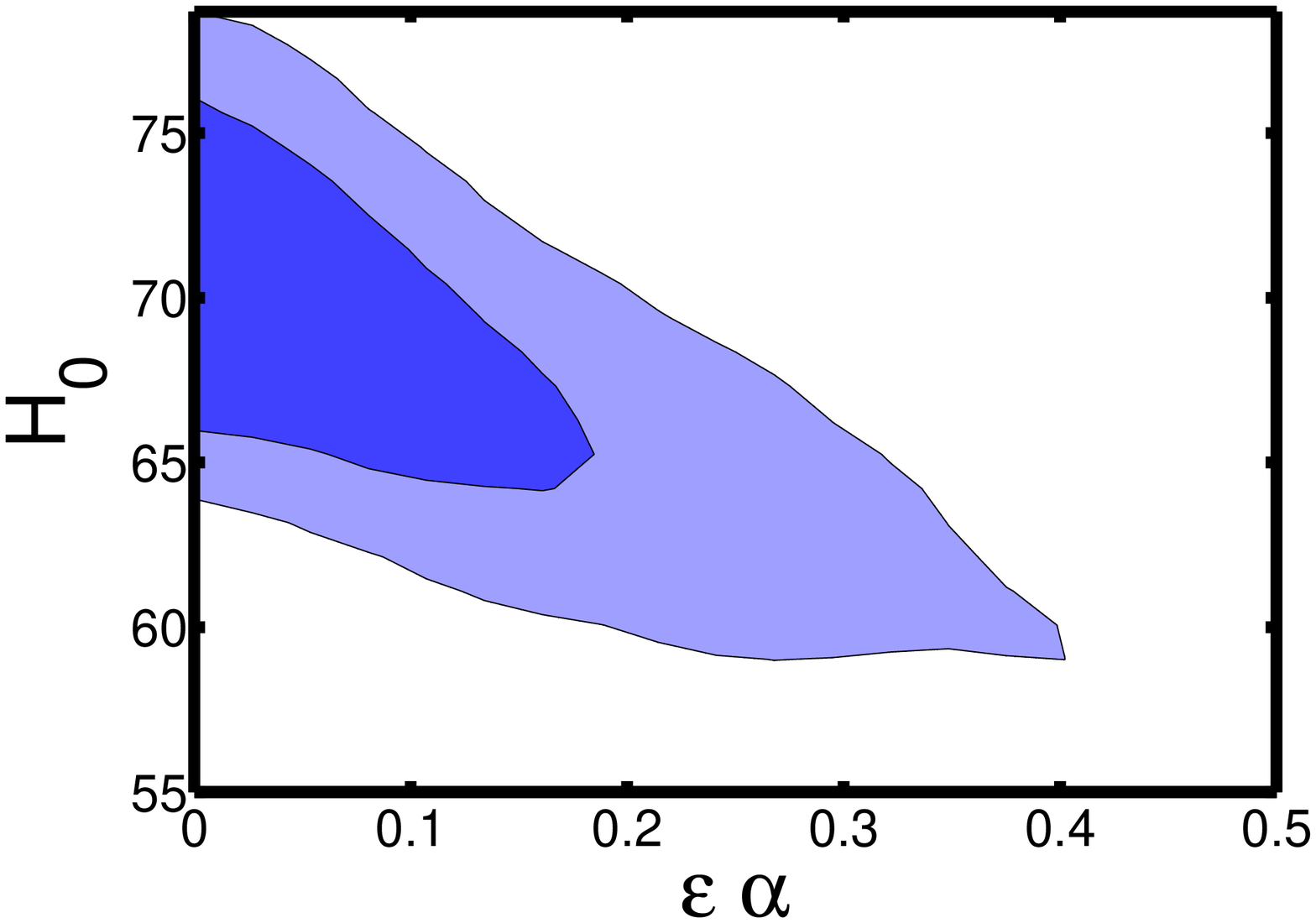}
\includegraphics[width=2.5in]{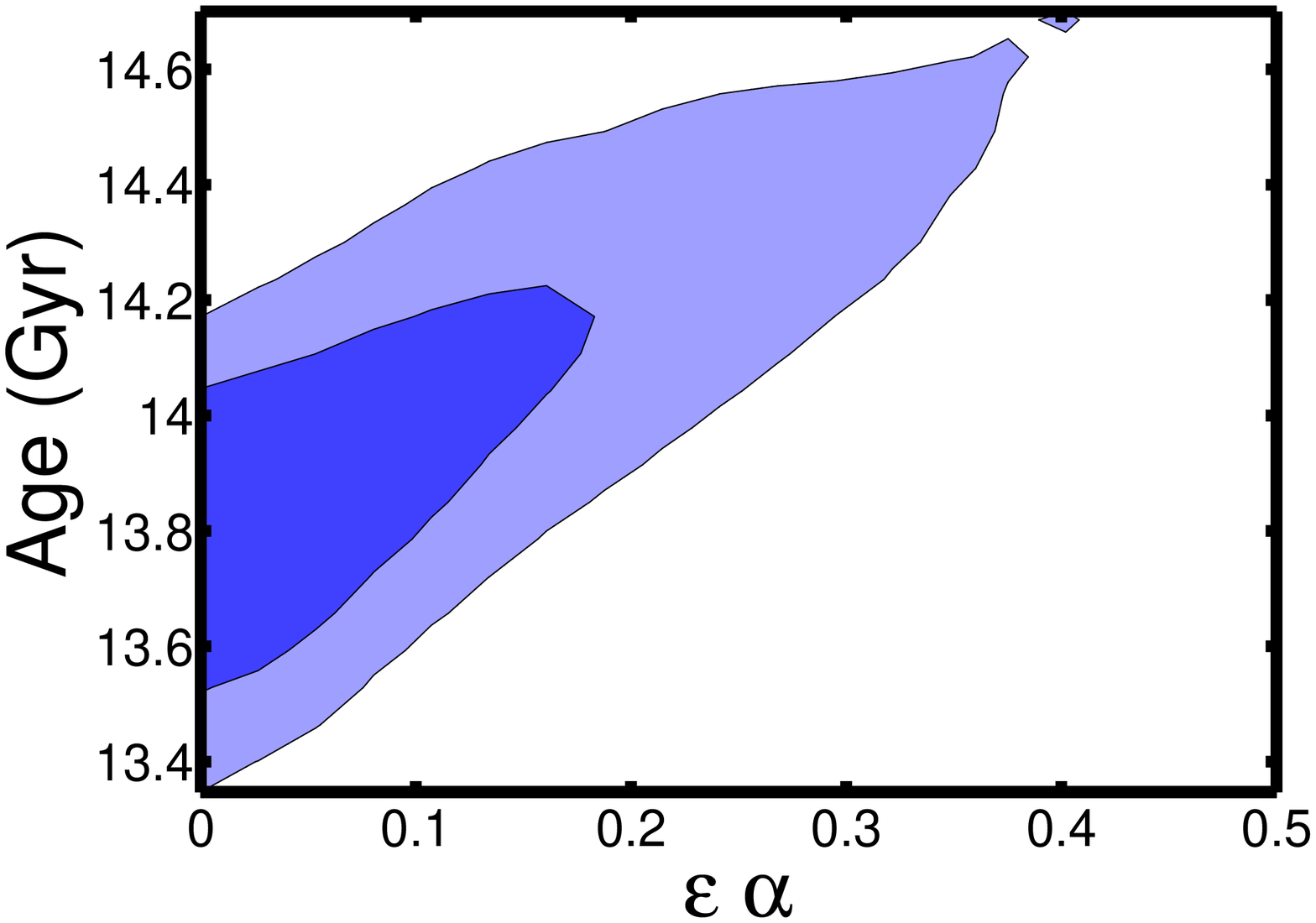}
\caption{Likelihood contours at  $68 \%$ and 
$95 \%$ c.l. on the $H_0-\epsilon_{\alpha}$,
(top panel) and $t_0-\epsilon_{\alpha}$
plane for WMAP5 fixing $\epsilon_i=0$ (bottom panel).
The Hubble constant is in units of $km/s/Mpc$.}
\label{fig4}
\end{center}
\end{figure}

The degeneracy with the spectral index can be 
easily understood: delayed recombination
suppresses the amplitude of the acoustic peaks in the CMB angular temperature anisotropy 
in a very similar way to an increase in the optical depth
\cite{Seager00}. 
This effect can be compensated by an increase
in $n_S$. Ionizing reionization also produces significant large
angular scale polarization, and WMAP EE and TE constraints 
help to break the $n_s-\epsilon_i$ degeneracy. However the same is not
true for resonance recombination 
\cite{bms2}, and so the $n_s-\epsilon_\alpha$ degeneracy remains present despite improved polarization measurements. 

Another effect of delayed recombination on the CMB
spectrum, through the presence of resonance radiation, is a shift of the acoustic peak spectrum towards
larger angular scales. An increase in $\epsilon_{\alpha}$ 
shifts the epoch of recombination towards smaller $z_*$.
This increases the size of the acoustic horizon at 
recombination and therefore shifts the peak towards smaller
$\ell$'s (see \cite{Seager00},\cite{bms2}). In a flat universe, this effect can 
be counter-balanced by a decrease of the Hubble constant.
In Figure \ref{fig4} (Top Panel) we plot the likelihood contours
from WMAP5 on the $H_0-\epsilon_{\alpha}$ plane. A degeneracy with
the Hubble parameter is evident and the analysis yields
$H_0=68.1_{-8.6}^{+7.2}$ at $95 \%$ c.l. that should compared with the
value $H_0=71.9_{-5.4}^{+5.2}$ at $95 \%$ c.l. assuming standard
recombination. Since smaller values of the Hubble constant 
are in agreement with larger $\epsilon_{\alpha}$ it is easy
to predict a correlation with the current age of the universe $t_0$.

In the bottom panel of Figure \ref{fig4} we plot
the likelihood contours in the $t_0-\epsilon_{\alpha}$ plane.
As we see, larger values of $\epsilon_{\alpha}$ are indeed in 
agreement with larger $t_0$. We find from WMAP5 that
$t_0=14.0_{-0.4}^{+0.6}$ Gyrs to be compared with
the constraint $t_0=13.7_{-0.3}^{+0.3}$ under standard recombination.
Current CMB age constraints can, therefore, 
be underestimated if one assumes a standard recombination scheme.

\begin{figure}[t]
\begin{center}
\includegraphics[width=2.5in]{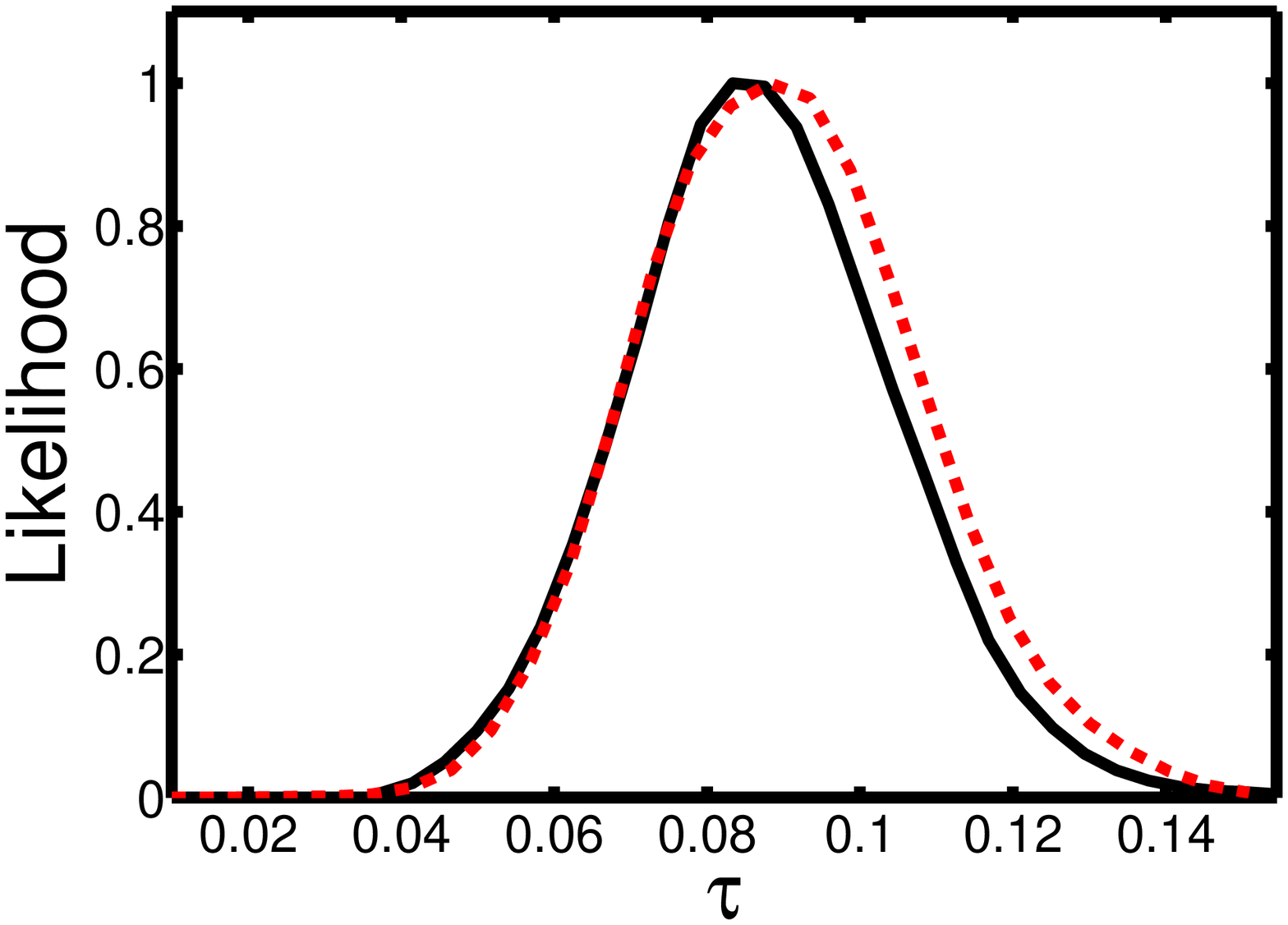}
\includegraphics[width=2.6in]{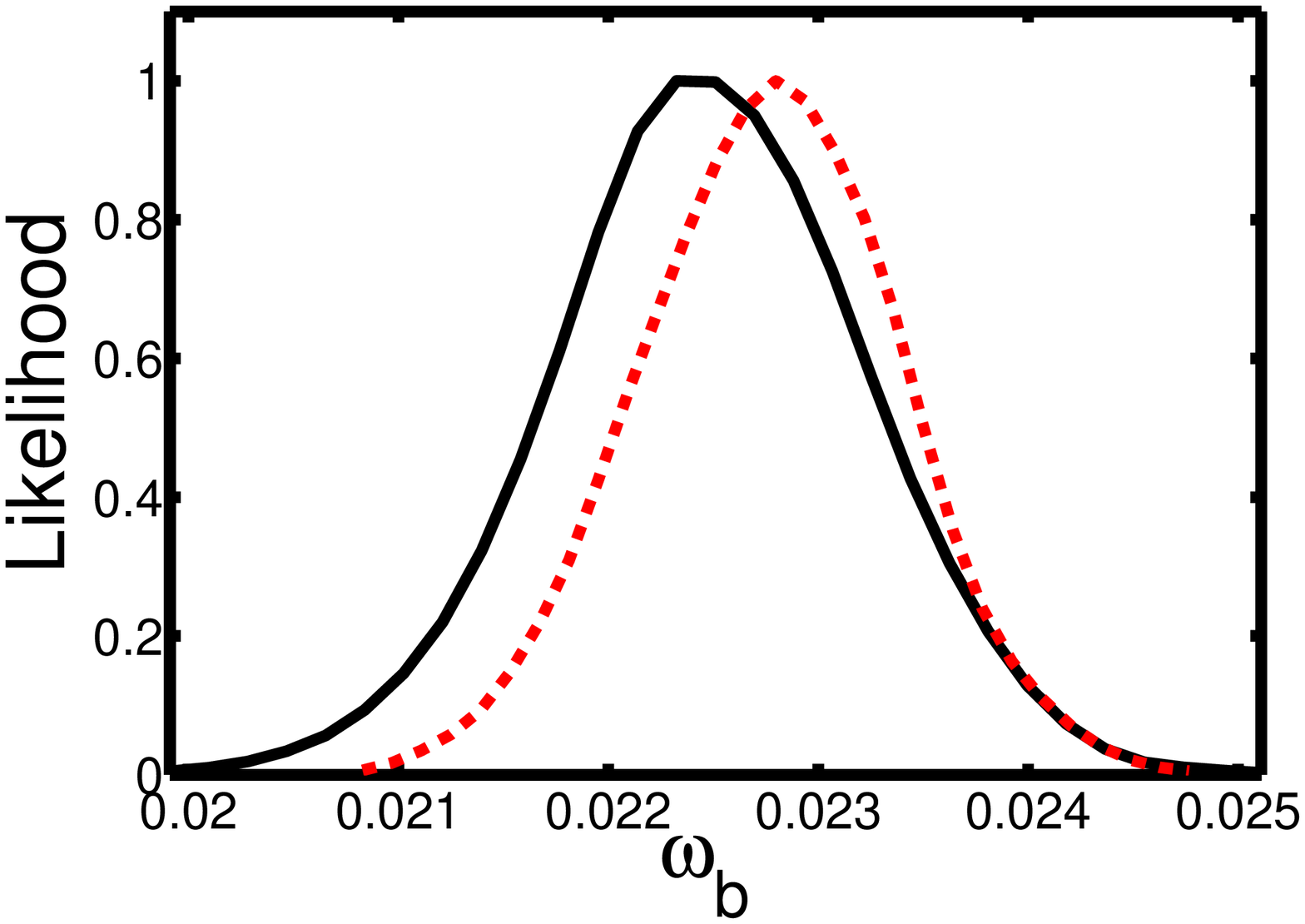}
\includegraphics[width=2.5in]{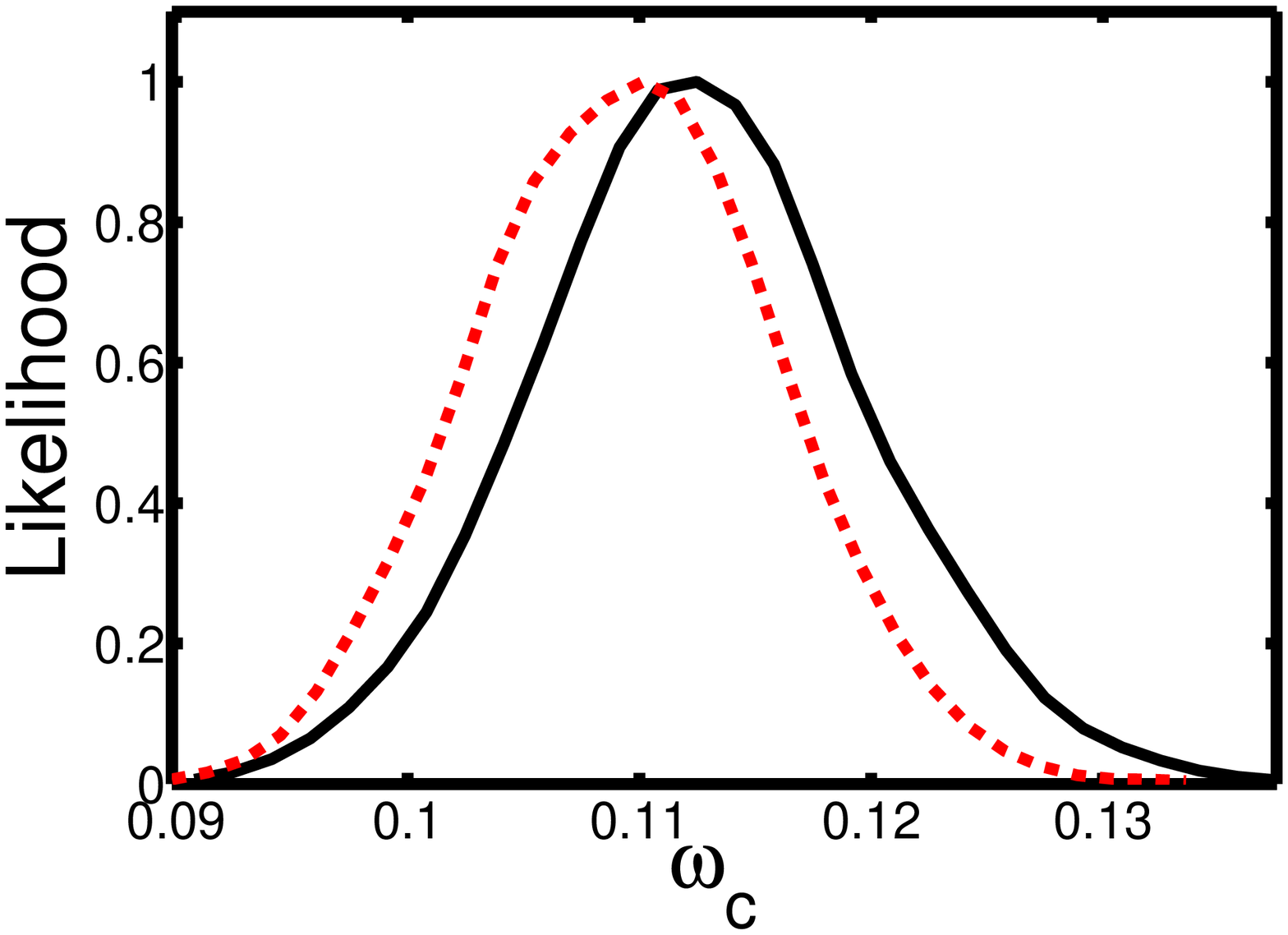}
\caption{Likelihood distribution function from 
WMAP5 on the optical depth (top panel),
the baryon density (center panel) and the
cold dark matter density (bottom panel). The
red dashed line are the results assuming standard 
recombination, the solid black line are under
the hypothesis of delayed recombination.}
\label{fig5}
\end{center}
\end{figure}

The impact of delayed recombination on the remaining
parameters is shown in
 Figure \ref{fig5}  where we plot the 1-D likelihood functions for 
the optical depth $\tau$ and the baryon and cold dark
matter densities $\omega_b$ and $\omega_c$ derived from the
WMAP5 data with and without standard recombination.
While the changes in the optical depth
are minimal since this parameter is mainly fixed
by large scale polarization data, delayed recombination may
slightly bias the current constraints towards smaller 
$\omega_b$ and larger $\omega_c$ respect to the standard case.

Another parameter worthy of study, and
on which the CMB usually provides clear and uncorrelated
constraints, is the shift parameter, $\cal{R}$,  
(see eg. \cite{corasaniti}) which defines 
 the angular scale of the acoustic oscillations.
The shift parameter is generally considered a 
 simple, geometric way to implement information from CMB
in determining constraints from independent data sets such  
type Ia supernovae or galaxy clustering.
However this parameter is not directly measured by CMB observations but 
is essentially a byproduct of the cosmological 
parameter inference based on  CMB spectral data analysis, and most crucially the recombination history. Therefore,
using the shift $\cal{R}$ to constrain cosmological parameters in
combination with other datasets may suffer from model dependent
assumptions. It is subsequently important to
investigate the effects modified recombination can have on this key parameter.
We find WMAP5 with delayed recombination constrains
${\cal R}=1.737\pm0.028$  at $68 \%$ c.l., to be compared with 
${\cal R}=1.713\pm0.020$ at $68 \%$ c.l. for standard recombination.
Such a $1-\sigma$ shift in $\cal{R}$ should not, however, alter in a significant way 
the current constraints on dark energy parameters derived 
from combined analyses assuming the standard recombination.

As expected, modified recombination strongly 
affects the constraints on the redshift 
of recombination, $z_*$. According to the latest WMAP5 results, 
the redshift of recombination $z_*$ is $z_*=1090.51\pm0.95$
with a better than $0.1 \%$ precision. However this impressive
result is mainly driven by the assumption 
of standard recombination itself.
With delayed recombination,
we found $z_*=1078.2\pm10.9$ at $68 \%$ c.l.
from WMAP5 alone and $z_*=1084.2\pm9.9$ with WMAP5+ACBAR, so that 
the uncertainty in the measurement of $z_*$ is increased by an order of magnitude.

\begin{figure}[t]
\begin{center}
\includegraphics[width=3.2in]{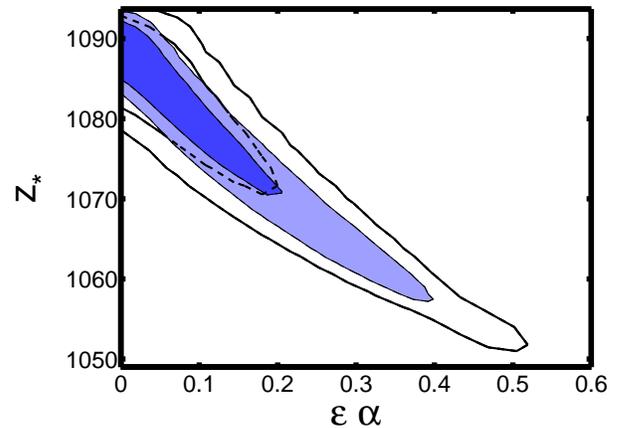}
\caption{Constraints on the $z_*$-$\epsilon_{\alpha}$
plane from WMAP5 (empty contours) and WMAP5+ACBAR (filled contours).
A degeneracy along the relation
 $z_*(\epsilon_{\alpha})=z_*(0)(1+3\epsilon_{\alpha})^{-0.042}$
is evident.}
\label{fig6}
\end{center}
\end{figure}

In Figure \ref{fig6} we plot the 2-D likelihood contours for
$z_*$ and $\epsilon_{\alpha}$. A degeneracy following the
relation $z_*(\epsilon_{\alpha}) \simeq z_*(0)(1+3\epsilon_{\alpha})^{-0.042}$ 
(see \cite{Seager00}) is clearly evident and explains why
the constraints on $z_*$ are much broader in the presence of delayed 
recombination.

\subsection{Modified recombination with curvature}
As mentioned in the previous section,
 delayed recombination decreases $z_*$, increasing
 the size of the sound horizon at recombination. 
This shifts the position of the
acoustic peaks in the CMB spectrum as 
$\ell \sim (z_*/z_*^{standard})^{1/2}$.  We therefore 
expect a degeneracy between $\epsilon_{\alpha}$ and 
the curvature $\omega_k$: namely a lower recombination redshift
shifts the CMB peaks to larger angular scales (smaller $\ell$'s) 
yielding open models more consistent with the data (see \cite{Seager00}).
However, as already pointed out in 
\cite{Seager00}, measurements of the multiple peaks in the angular spectrum 
should break this degeneracy. 
In this section we perform an analysis of delayed recombination in 
curved models and check if a flat universe is still consistent with
the data and test to what extent open models can be in agreement
with observations. Including curvature, we found that the constraints
on WMAP5+ACBAR are slightly relaxed with 
$\epsilon_\alpha < 0.39$ and $\epsilon_i < 0.058$
at $95 \%$ c.l.
However the effect on curvature from delayed recombination is 
small. Marginalizing over the recombination parameters we get 
$\omega_k=-0.033_{-0.100}^{+0.058}$ (WMAP5+ACBAR) 
at $95 \%$ c.l..  to be compared with the constraint
 $\omega_k=-0.046_{-0.094}^{+0.062}$ obtained using the same
datasets and priors but with standard recombination.
The shifts towards open models due to delayed recombination
 is therefore at the level of few percent.

\begin{figure}[t]
\begin{center}
\includegraphics[width=3.2in]{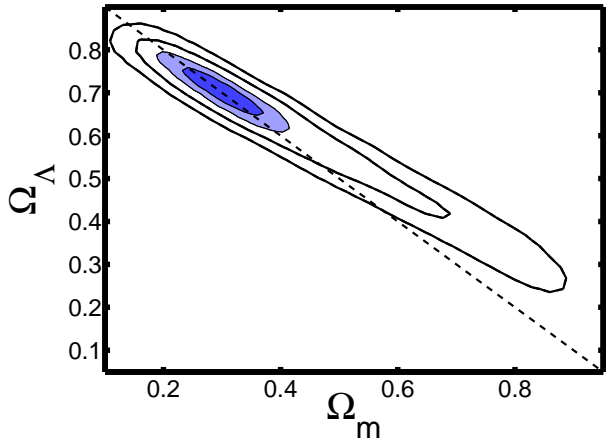}
\caption{Constraints on the $\Omega_M$-$\Omega_{\Lambda}$
plane  in the presence of delayed recombination from WMAP5+ACBAR (empty contours) and after adding
an external prior $\Omega_M=0.30 \pm 0.05$ (filled contours).}
\label{fig7}
\end{center}
\end{figure}

When external priors are added,  we see 
little variation in the current constraints on $\Omega_{\Lambda}$ in
a curved universe. We show this in Figure \ref{fig7}
where we plot the usual constraints on the $\Omega_M$-$\Omega_{\Lambda}$
but including delayed recombination. The changes are minimal
respect to standard recombination. Adding a prior on the matter
parameter $\Omega_M=0.30 \pm 0.05$ yields a constraint
 $\Omega_{\Lambda}=0.70 \pm 0.04$ 
(to be compared with  $\Omega_{\Lambda}=0.712 \pm 0.037$ with standard
recombination) still suggesting the presence of
a cosmological constant at high significance.

\subsection{Modified recombination and massive neutrinos}
 
Cosmological neutrinos have a relevant impact on
cosmology since they change the expansion history of the universe
and affect the growth of perturbations. CMB anisotropies
can constrain neutrino masses indirectly since variations in the
gravitational potential change the shape of the small scale
CMB angular spectrum.

The recent WMAP5 data have provided an upper limit to
the sum of neutrino masses of $\Sigma m_{\nu} < 1.3 eV$ at
$95 \%$ c.l. (\cite{wmap5komatsu}). Since this is the best upper limit
current available on absolute neutrino masses it is
certainly timely to investigate how this boud could change
when delayed recombination is included.
We have found that from WMAP5+ACBAR the bound on 
neutrino masses is $\Sigma m_{\nu} < 1.26 eV$ at $95 \%$ c.l.
under delayed recombination larger than
$\Sigma m_{\nu} < 1.13 eV$ obtained under standard recombination
with the same dataset and priors. We can therefore conclude that
delayed recombination increases the current bound by $\sim 10\%$.

\subsection{Forecasts for the Planck Surveyor}
In this section, we forecast the future constraints achievable 
on modified recombination from the
Planck satellite experiment. We have 
constrained cosmic parameters assuming 
simulated Planck mock data with a fiducial model
given by the best fit WMAP5 model (with standard recombination) 
and experimental noise described by
\begin{equation}
N_{\ell} = \left(\frac{w^{-1/2}}{\mu{\rm K\mbox{-}rad}}\right)^2 
\exp\left[\frac{\ell(\ell+1)(\theta_{\rm FWHM}/{\rm rad})^2}{8\ln 2}\right],
\end{equation}
with $w^{-1/2}=63 \mu K$ as the temperature noise level
(we consider a factor $\sqrt{2}$ larger for polarization 
noise) and  $\theta_{\rm FWHM}=7'$ for the beam size.
We take $f_{sky}=0.65$ as sky coverage.

With this configuration, we found the following constraints
from Planck: $\epsilon_{\alpha}<0.01$ and  
$\epsilon_{i}<0.0005$ at $95\%$ c.l., representing more than an order of magnitude in improvement in $\epsilon_\alpha$ and two order of magnitude improvement in $\epsilon_i$, respectively, in comparison to current WMAP5+ACBAR constraints. 

\begin{figure}[t]
\begin{center}
\includegraphics[width=2.5in]{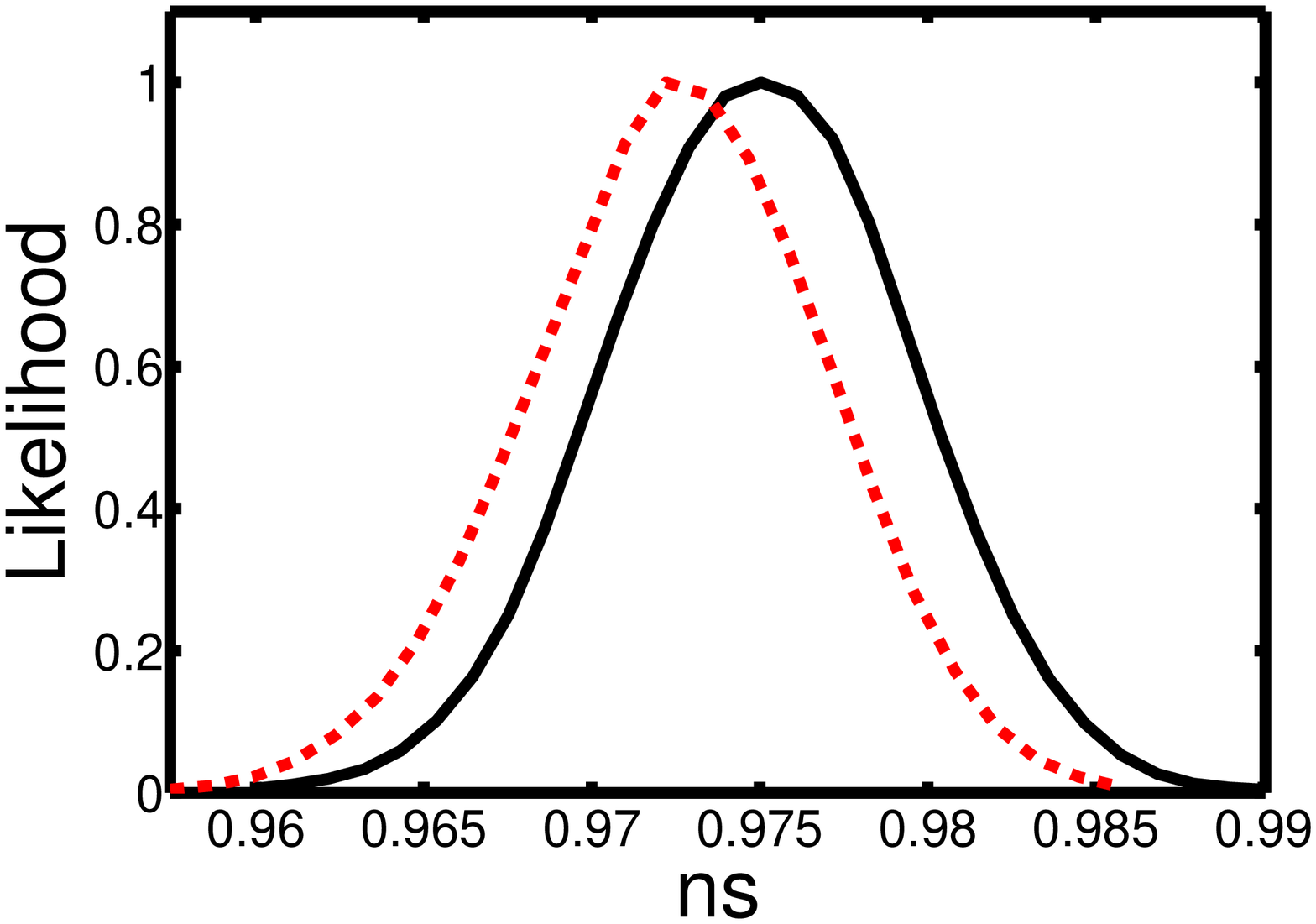}
\includegraphics[width=2.6in]{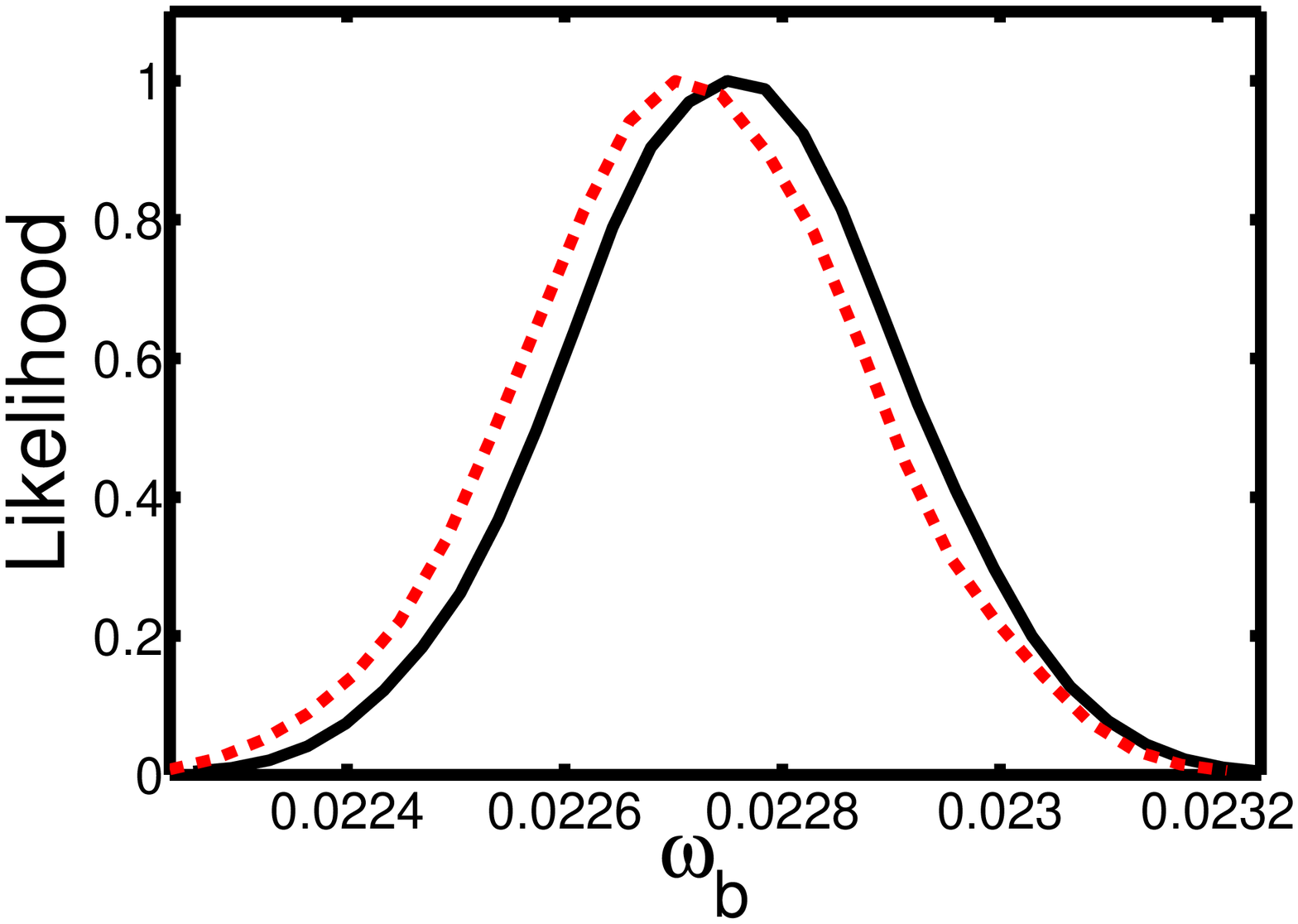}
\includegraphics[width=2.5in]{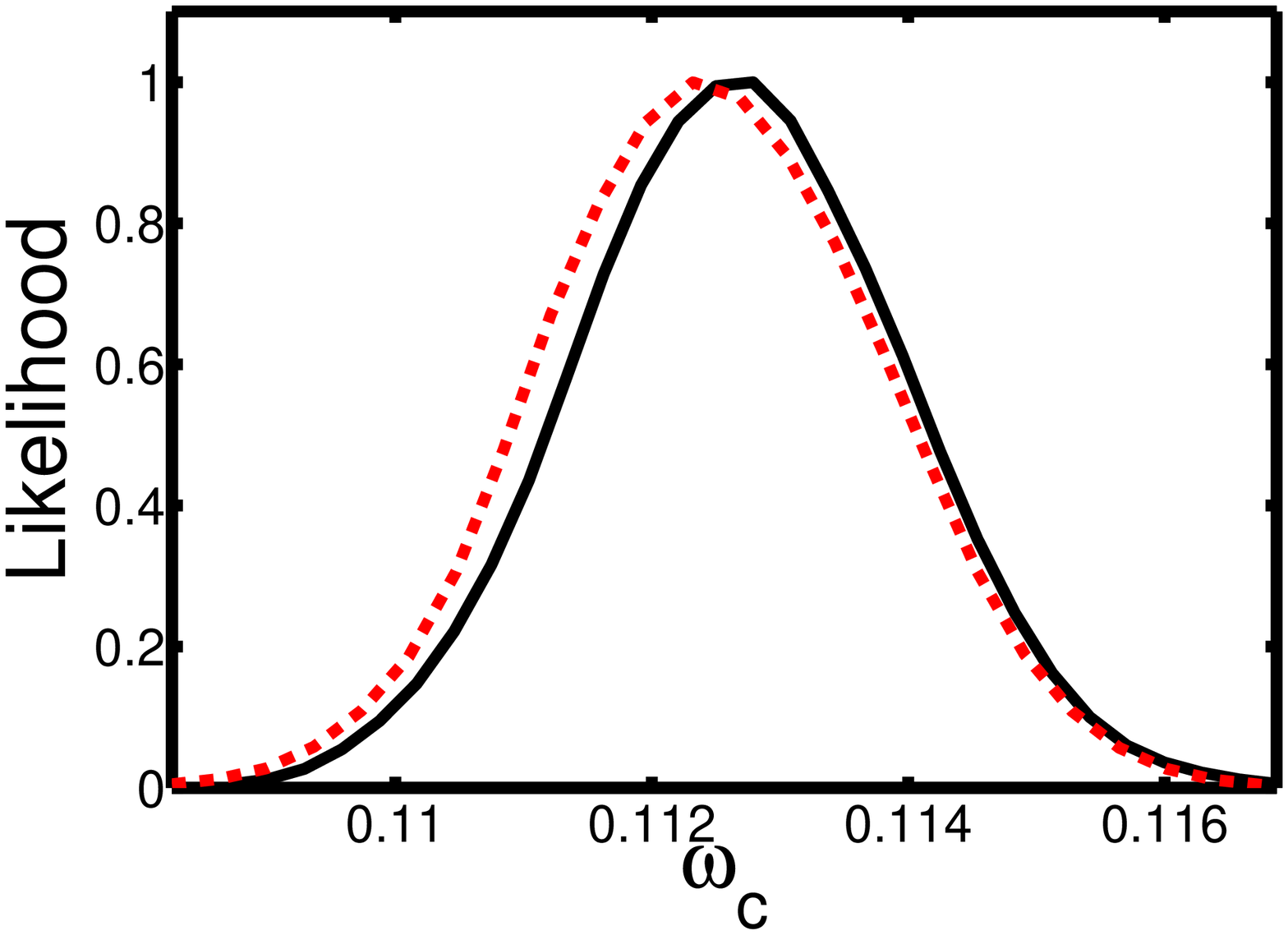}
\caption{Likelihood distribution function from 
a Planck-like satellite experiment (see text for the experimental
configuration) on the scalar spectral index (top panel),
the baryon density (center panel) and the
cold dark matter density (bottom panel). The
red dashed line are the results assuming standard 
recombination, the solid black line are under
the hypothesis of delayed recombination.}
\label{fig8}
\end{center}
\end{figure}

\begin{figure}[t]
\begin{center}
\includegraphics[width=2.5in]{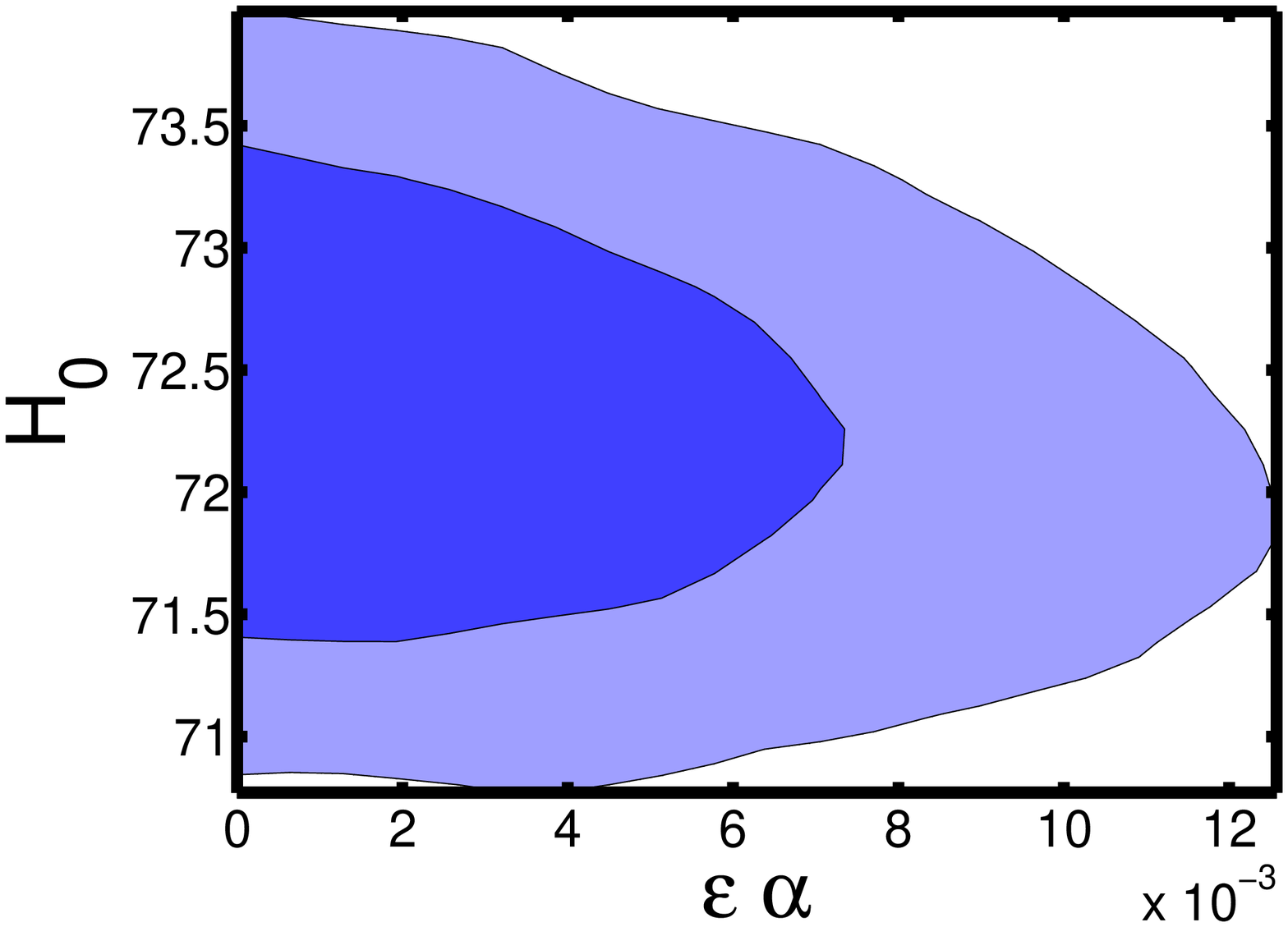}
\includegraphics[width=2.6in]{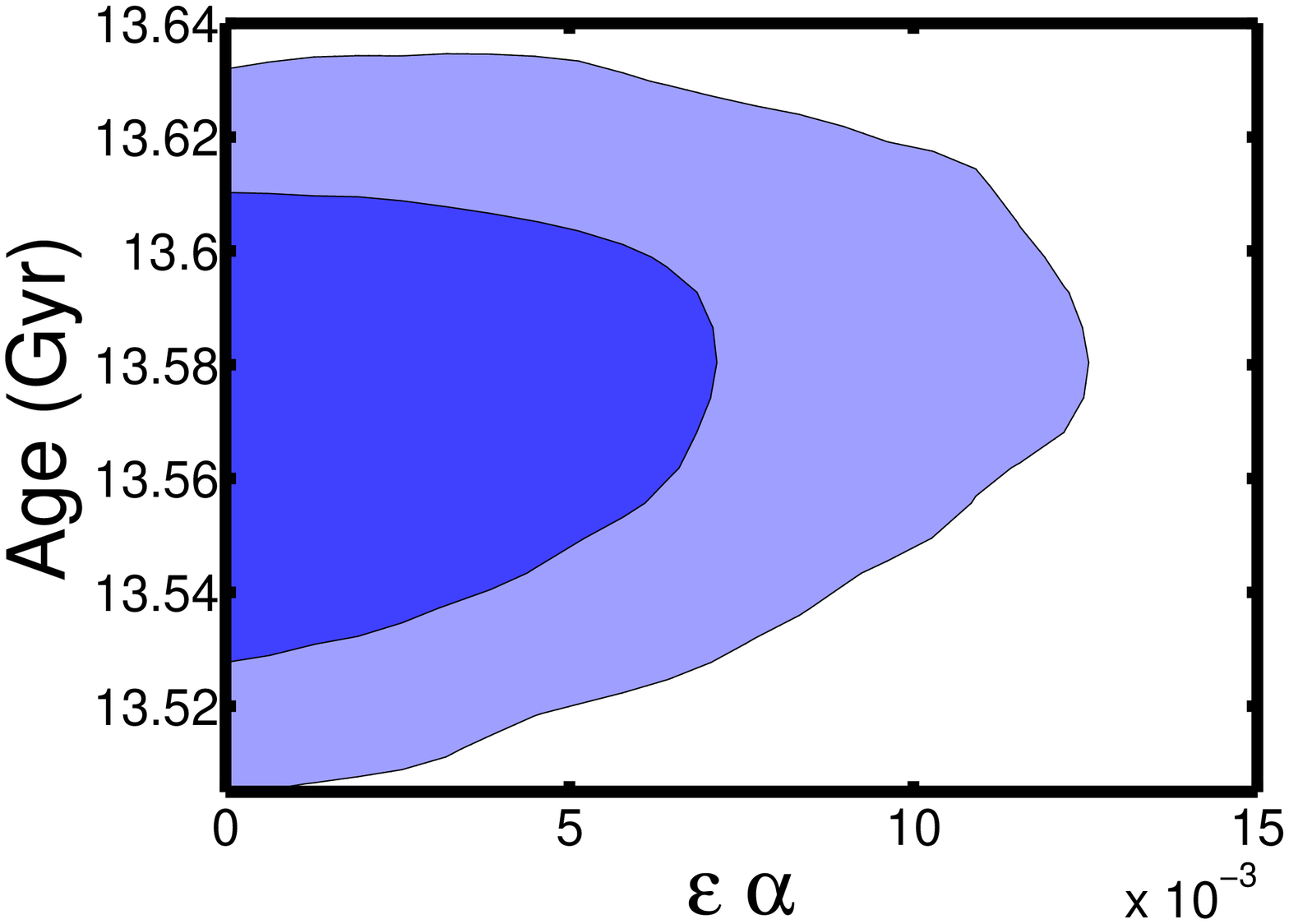}
\caption{2D likelihood contour plots at $68 \%$ and
$95 \%$ c.l. from a Planck-like satellite experiment 
(see text for the experimental configuration) for the Hubble parameter 
(top Panel) and age (Bottom Panel) versus the rcombination
parameter $\epsilon_{\alpha}$.}
\label{fig9}
\end{center}
\end{figure}

In Figures \ref{fig8} and  \ref{fig9} we show the effects
 of delayed recombination on constraining
cosmological parameters from Planck. 
As we can see, the precision small scale temperature measurements of 
Planck markedly reduce the $n_s-\tau-\epsilon_i/\epsilon_\alpha$
 degeneracies. $\epsilon_i$ is 
further constrained by small scale polarization measurements.
 Subsequently a modified recombination 
scheme should result in values of the spectral index $n_s$ being only slightly skewed towards more positive values in 
comparison to the standard scheme, and the Planck satellite will therefore be clearly able to discriminate 
$n_s$ from a HZ spectrum with high significance 
even in the case of delayed recombination. The significantly larger uncertainties  in $t_0$ and $H_0$ currently introduced if delayed, rather than standard, recombination is considered with the WMAP5 +ACBAR data would be significantly curtailed with Planck measurements.

\section{Conclusions}
\label{sec4}

In this paper, we have updated the upper bounds that can be placed on the contribution 
of extra Ly-$\alpha$ and ionizing photon-producing sources in light of
the new WMAP and ACBAR data.  
We have found that, adopting a simple parametrization using constant,
effective values for $\epsilon_{\alpha}$ and $\epsilon_{i}$,
the WMAP5 data constraints $\epsilon_{\alpha} < 0.31$ and 
$\epsilon_{i}<0.053$ at the $95\%$ level.

We have studied the implications of delayed recombination on several
current cosmological constraints derived from CMB anisotropies.
While we found no relevant changes on current CMB estimates 
of curvature, optical depth, baryon and cold dark matter densities, delayed recombination
proves to have a significant impact on current constraints for
inflationary parameters. Current conclusions for theoretical
inflationary models, 
motivated by evidence for a deviation from scale invariance from
recent CMB data 
in the standard recombination scheme, would have to be relaxed when
delayed recombination 
is considered, since the inflationary spectral index of scalar 
perturbations is in complete agreement with a scale invariant HZ spectrum. 

The delayed recombination also has an impact on estimates of the shift
factor, $\cal R$, and recombination redshift, 
$z_*$, currently used in estimating angular diameter distance and baryon acoustic oscillation measures of the cosmic expansion history.
Moreover, constraints on particle physics parameters like
the neutrino mass can also be relaxed when non-standard recombination
is considered.

Physically motivated models for non-standard recombination, like those 
based on primordial black hole or super heavy dark matter decay, are 
possible and provide a good fit to the current data. 
Future observations in both temperature and polarization, 
as those expected from the Planck satellite will be needed if we are to more stringently test and reduce the 
dependency of other cosmological parameters on delayed recombination models.

\medskip 
\textbf{Acknowledgments} 

This research has been supported by ASI contract I/016/07/0 "COFIS".
RB's work is supported by NASA ATP grant NNX08AH27G, NSF grants AST-0607018 and PHY-0555216 and Research Corporation.
We wish to thank Wayne Hu, Pavel Naselsky and Jim Peebles for useful 
comments.



\begin{thebibliography}{99}

\bibitem{wmap5cosm}
  G.~Hinshaw {\it et al.}  [WMAP Collaboration],
  arXiv:0803.0732 [astro-ph].

\bibitem{wmap5komatsu}
  E.~Komatsu {\it et al.},
  arXiv:0803.0547 [astro-ph].

\bibitem{acbar}
  C.~L.~Reichardt {\it et al.},
  arXiv:0801.1491 [astro-ph].

  
\bibitem{fogli}
  G.~L.~Fogli {\it et al.},
  arXiv:0805.2517 [hep-ph].

\bibitem{fdebe08}
  F.~De Bernardis, L.~Pagano, P.~Serra, A.~Melchiorri and A.~Cooray,
  JCAP {\bf 0806} (2008) 013
  [arXiv:0804.1925 [astro-ph]].

\bibitem{kinney08}
  W.~H.~Kinney, E.~W.~Kolb, A.~Melchiorri and A.~Riotto,
  arXiv:0805.2966 [astro-ph].

\bibitem{Peebles68} P.J.E. Peebles, \ApJ {\bf 153} 1 (1968).
  
\bibitem{Zeldovich68} Ya. B. Zel'dovich, V.G. Kurt, R.A. Sunyaev, {\it Zh. Eksp. Teoret. Fiz} {\bf 55} 278(1968), English translation, {\it Sov. Phys. JETP.} {\bf 28} 146 (1969).

\bibitem{Seager99} S. Seager, D.D. Sasselov, \& D. Scott, \ApJ {\bf 523} 1 (1999), astro-ph/9909275.
\bibitem{Hu95} W. Hu, D. Scott, N. Sugiyama, \& M. White, \PR D {\bf 52} 5498 (1998).

\bibitem{Seljak03}  U. Seljak, N. Sugiyama, M. White, M. Zaldarriagaastro-ph/0306052

\bibitem{hirata}
  E.~R.~Switzer and C.~M.~Hirata,
  Phys.\ Rev.\  D {\bf 77} (2008) 083006
  [arXiv:astro-ph/0702143];
  C.~M.~Hirata and E.~R.~Switzer,
  Phys.\ Rev.\  D {\bf 77} (2008) 083007
  [arXiv:astro-ph/0702144].
  
  \bibitem{Hannestad01} S. Hannestad \& R. J.  Scherrer, \PR D {\bf 63}, 083001 (2001)

\bibitem{Lewis:2006ym}
  A.~Lewis, J.~Weller and R.~Battye,
  Mon.\ Not.\ Roy.\ Astron.\ Soc.\  {\bf 373}, 561 (2006)
  [arXiv:astro-ph/0606552].
  
  \bibitem{alpha} S. Hannestad, \PR D {\bf 60}, 023515 (1999); 
M. Kaplinghat, R. J. Scherrer \& M S Turner, \PR D {\bf 60},023516 (1999); 
P. P. Avelino et al.,PRD {\bf 62}, 123508 (2000); 
R. Battye, R. Crittenden \& J. Weller, \PR D {\bf 63}, 043505 (2001); 
P.~P.~Avelino {\it et al.},
\PR D {\bf 64} (2001) 103505
[arXiv:astro-ph/0102144]
Landau, Harari \& Zaldarriaga, \PR D {\bf 63}, 083505 (2001).
  C.~J.~A.~Martins, A.~Melchiorri, G.~Rocha, R.~Trotta, P.~P.~Avelino and P.~Viana,
  Phys.\ Lett.\ B {\bf 585}, 29 (2004)
  [arXiv:astro-ph/0302295].
 G.~Rocha, R.~Trotta, C.~J.~A.~Martins, A.~Melchiorri, P.~P.~Avelino, R.~Bean and P.~T.~P.~Viana,
  Mon.\ Not.\ Roy.\ Astron.\ Soc.\  {\bf 352} (2004) 20
  [arXiv:astro-ph/0309211].
 C.~J.~A.~Martins, A.~Melchiorri, R.~Trotta, R.~Bean, G.~Rocha, P.~P.~Avelino and P.~T.~P.~Viana,
  Phys.\ Rev.\  D {\bf 66} (2002) 023505
  [arXiv:astro-ph/0203149].


\bibitem{zalzahn}
  O.~Zahn and M.~Zaldarriaga,
  Phys.\ Rev.\  D {\bf 67} (2003) 063002
  [arXiv:astro-ph/0212360].

\bibitem{Seager00} P.J.E. Peebles, S. Seager, W. Hu, \ApJ {\bf 539} L1 (2000), astro-ph/0004389. 

\bibitem{Naselsky02} P.D. Naselsky, I.D. Novikov {\it MNRAS} {\bf 334} 137 (2002), astro-ph/0112247

\bibitem{Dorosh02} A.G. Doroshkevich, I.P. Naselsky, P.D. Naselsky, I.D. Novikov, \ApJ {\bf 586} 709 (2002), astro-ph/0208114.


\bibitem{Sarkar:1983} S.~Sarkar and A.~Cooper, Phys. Lett. B., {\bf 148}, 347 (1983)

\bibitem{Scott:1991}, D. Scott ,M.~J.~ Rees  \& D.~W.~ Sciama , A \& A, {\bf 250}, 295, (1991)

\bibitem{Ellis:1992} J. Ellis, G. Gelmini , J. Lopez , D. Nanopoulos, \& S. Sarkar , Nucl.Phys.B. {\bf 373} 399 (1992).

\bibitem{Adams:1998} A. J. Adams J.A., S. Sarkar \& D.W. Sciama , MNRAS, {\bf 301}, 210 (1998)

\bibitem{Doroshkevich:2002ff}
  A.~G.~Doroshkevich and P.~D.~Naselsky,
  Phys.\ Rev.\ D {\bf 65}, 123517 (2002)
  [arXiv:astro-ph/0201212].

\bibitem{Naselsky:2003zj}
  P.~D.~Naselsky and L.~Y.~Chiang,
  Phys.\ Rev.\ D {\bf 69}, 123518 (2004)
  [arXiv:astro-ph/0312168].

\bibitem{Zhang:2006fr}
  L.~Zhang, X.~L.~Chen, Y.~A.~Lei and Z.~G.~Si,
  Phys.\ Rev.\ D {\bf 74}, 103519 (2006)
  [arXiv:astro-ph/0603425].

\bibitem{Pierpaoli:2003rz}
  E.~Pierpaoli,
  Phys.\ Rev.\ Lett.\  {\bf 92}, 031301 (2004)
  [arXiv:astro-ph/0310375].
  
\bibitem{Chen:2003gz}
  X.~L.~Chen and M.~Kamionkowski,
  Phys.\ Rev.\ D {\bf 70}, 043502 (2004)
  [arXiv:astro-ph/0310473].

\bibitem{Padmanabhan:2005es}
  N.~Padmanabhan and D.~P.~Finkbeiner,
  Phys.\ Rev.\  D {\bf 72}, 023508 (2005)
  [arXiv:astro-ph/0503486].
  
\bibitem{Mapelli:2006ej}
  M.~Mapelli, A.~Ferrara and E.~Pierpaoli,
  Mon.\ Not.\ Roy.\ Astron.\ Soc.\  {\bf 369}, 1719 (2006)
  [arXiv:astro-ph/0603237].
    
    \bibitem{Lewis:2006ma}
A. Lewis,
arXiv:astro-ph/0603753.

\bibitem{Naselsky87} P.D. Naselsky \& A.G Polnarev, {\it Sov. Astron. Lett.} {\bf 13} 67 (1987)

\bibitem{bms}
  R.~Bean, A.~Melchiorri and J.~Silk,
  Phys.\ Rev.\ D {\bf 68} (2003) 083501
  [arXiv:astro-ph/0306357].

\bibitem{bms2}
  R.~Bean, A.~Melchiorri and J.~Silk,
  Phys.\ Rev.\  D {\bf 75} (2007) 063505
  [arXiv:astro-ph/0701224].

\bibitem{naso}
  J.~Kim and P.~Naselsky,
  arXiv:0802.4005 [astro-ph].

\bibitem{petruta}
  P.~Stefanescu,
  New Astron.\  {\bf 12} (2007) 635
  [arXiv:0707.0190 [astro-ph]].

\bibitem{cinesi}
  L.~Zhang, X.~Chen, M.~Kamionkowski, Z.~g.~Si and Z.~Zheng,
  Phys.\ Rev.\  D {\bf 76}, 061301 (2007)
  [arXiv:0704.2444 [astro-ph]].

\bibitem{Eisenstein:2005su}
  D.~J.~Eisenstein {\it et al.}  [SDSS Collaboration],
   ``Detection of the Baryon Acoustic Peak in the Large-Scale Correlation
  Astrophys.\ J.\  {\bf 633}, 560 (2005)
  [arXiv:astro-ph/0501171].

\bibitem{Percival:2007yw}
  W.~J.~Percival, S.~Cole, D.~J.~Eisenstein, R.~C.~Nichol, J.~A.~Peacock, A.~C.~Pope and A.~S.~Szalay,
  Mon.\ Not.\ Roy.\ Astron.\ Soc.\  {\bf 381}, 1053 (2007)
  [arXiv:0705.3323 [astro-ph]].

\bibitem{efstathiou} Efstathiou, G. \& Bond J. R. 
1999, MNRAS, 304, 75


\bibitem{corasaniti}
  P.~S.~Corasaniti and A.~Melchiorri,
  Phys.\ Rev.\  D {\bf 77}, 103507 (2008)
  [arXiv:0711.4119 [astro-ph]].

\bibitem{Elgaroy:2007bv}
  O.~Elgaroy and T.~Multamaki,
  arXiv:astro-ph/0702343.

\bibitem{Lewis:2002ah}
A. Lewis and S. Bridle,
Phys.\ Rev.\ D {\bf 66}, 103511 (2002) (Available from 
\texttt{http://cosmologist.info}.)

\end{thebibliography}
\end{document}